\documentclass[11pt]{article}
\usepackage{amssymb}
\usepackage{amsmath}
\usepackage[dvips]{epsfig}
\begin{document}

\title{Shape resonances in nested diffraction gratings}
\author{Angela N. Fantino$^{1}$, Susana I. Grosz$^{2}$ and Diana C. Skigin$^{1}$, \\
{\em Grupo de Electromagnetismo Aplicado,}\\
{\em Departamento de F\'{\i}sica,}\\
{\em Facultad de Ciencias Exactas y Naturales, }\\
{\em Universidad de Buenos Aires, }\\
{\em Ciudad Universitaria, Pabell\'{o}n I, }\\
{\em C1428EHA Buenos Aires, Argentina}\\
{\em dcs@df.uba.ar}}
\date{}
\maketitle

\setcounter{footnote}{1} 
\footnotetext{Member of CONICET}
\setcounter{footnote}{2} 
\footnotetext{Ciclo B\'asico Com\'un, Universidad de Buenos Aires}

\baselineskip 4.5ex

\section*{Abstract}

The diffraction problem of a plane wave impinging on a grating formed by nested cavities
is solved by means of the modal method, for $s$ and $p$ polarization modes. The cavities
are formed by perfectly conducting sheets that describe rectangular profiles. The 
electromagnetic response of the grating is analyzed, paying particular attention to the
generation of resonances within the structure. The dependence of the resonances on the
geometrical parameters of the grating is studied, and results of far and near field 
are shown. The results are checked and compared 
with those available in the literature for certain limit cases.

{\bf Keywords:} surface-shape resonances, diffraction, gratings 
\newpage
\section{Introduction}

Current radio detection and ranging (RADAR) technologies employ highly concentrated
beams of electromagnetic energy to scan an area. The return echo is
then processed to detect all targets of interest \cite{Skolnik}. To prevent this kind of
detection, a target can either deflect the beam in a direction away from the observer, absorb
the energy of the incoming beam or alter the return echo by changing its frequency or
shape. Giving an aircraft textured skin that could interact with an incoming beam
adaptively would allow most shapes to be metamorphic to radar. Recent research based on
sub-wavelength, narrow-band, resonant, corrugated, high impedance surfaces have shown
promising results in this area \cite{Sievenpiper1,Sievenpiper2}. However, the speed,
bandwidth and adaptability of these surfaces are limited. To avoid the narrow-band
resonant behavior, lumped structures can be replaced by broadband antenna arrays
that add a controlled phase delay to the reflected signal allowing beam steering 
away from the observer \cite{Schaffner}. Fractal planar antennas 
\cite{Gianvittorio}-\cite{Loui} exhibit multi-band behavior and are small in electrical size, 
but their scattering behavior is not widely understood. Nested multilayer corrugated 
surfaces constitute an interesting alternative for this purpose. If successful, unwanted 
radiation due to finite ground planes can be minimized and the inter elemental coupling 
can be reduced, thus obtaining the desired effect. The study of the electromagnetic 
response of the two-layer periodic structure proposed in this paper is expected to provide
a first approach to the behavior of multilayer surfaces and their advantages as part of broadband 
antennas.

The resonant behavior of infinitely periodic gratings of rectangular profile 
was studied by many authors, in particular for $s$ polarization \cite{Hessel}-\cite{sad3}.
The excitation of surface shape resonances in structures comprising cavities have been lately
studied for particular profiles of the corrugation. The resonant features of an isolated cavity 
or groove of univalued and multivalued geometries have been investigated by means of different 
implementations of two basic approaches: the integral and the modal methods 
\cite{Ziolkowski}-\cite{diana14}. The results show that a strong intensification inside the cavity 
is found for certain wavelengths when its profile is bivalued, such as a slotted cylinder or a 
bottle-shaped groove \cite{Valencia}-\cite{diana13}. The effect produced by the surface shape
resonances on the response of an array of cavities has also been studied 
\cite{diana14}-\cite{Mata}. 

The modal method employed here to solve the diffraction problem from an infinite grating 
comprising rectangular grooves and bottle-shaped cavities, was first formulated to deal 
with simple geometries such as rectangular \cite{Andrewartha}, triangular \cite{Jovicevic} 
or semicircular \cite{Andrewartha1}. Later it was generalized for arbitrary
shapes of the grooves \cite{Andrewartha2,Li1}, and recently it was also applied to 
bottle-shaped cavities \cite{diana14}. The implementation of the modal method for perfectly
conducting structures results in a simple and efficient way of calculating the external and
the internal fields, without the need of any sophisticated algorithm. 

The purpose of this paper is to analyze the electromagnetic response of a two-layer perfectly-conducting
grating comprising grooves and cavities. Particular attention is paid to the resonant behavior of this structure and its effect in the field distribution inside and far from the corrugations. The study of
such gratings constitutes a first approach to the analysis of N-layer fractal gratings, which are expected to be useful for the control of the phase delay in the scattering response of
broadband microwave antennas.

In Sec. 2 we pose the problem and give the details of the modal method applied to the present structure,
for both fundamental polarization modes. In Sec. 3 we give numerical results for the reflected efficiency
as well as for the near field. We analyze the dependence of the resonances on the geometrical
parameters of the structure, and compare the results obtained in limit cases with those found in the
literature. Finally, some conclusions are summarized in Sec. 4.  

\section{Theoretical formulation}

We consider the diffraction of a plane wave by a two-layer nested grating. The structure is 
one-dimensional and infinitely periodic, and each period is formed by a bottle-shaped cavity and rectangular grooves, bounded by perfectly conducting sheets (see Fig. 1). 

\begin{figure}[h]
\begin{center}
\includegraphics[width=5in]{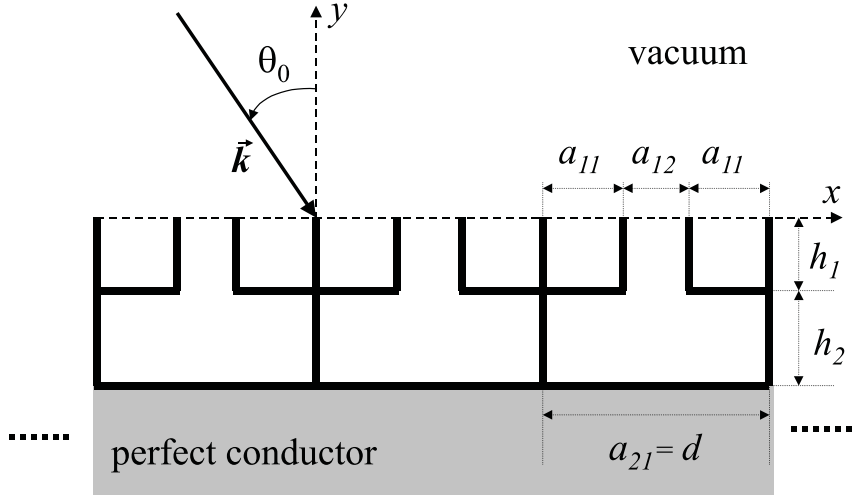}
\end{center}
\caption{Sketch of the nested diffraction grating.}
\end{figure}

The wavelength of the 
incoming wave is $\lambda$ and the angle of incidence is $\theta_0$. Since the structure and the 
fields are invariant under traslations in the $\hat{z}$ direction, the problem can be separated 
into two scalar
problems corresponding to the basic modes of polarization: $s$ (electric field parallel to the
rulings) and $p$ (magnetic field parallel to the rulings). In the upper layer, the period has
two rectangular grooves of width $a_{11}$ and height $h_1$ (zones 11 and 13, see Fig. 1), and 
a central neck of the same depth but of width $a_{12}$ (zone 12). The second layer is a 
rectangular cavity which occupies the whole period, i.e., its width is $d$ and its depth is 
$h_2$ (zone 21). The surrounding medium is vacuum. In what follows, both 
polarizations will be treated simoultaneously, denoting by $f$ the $z$-component of the electric 
field in the $s$-case and that of the magnetic field in the $p$-case.

In an homogeneous medium, the rectangular geometry of the structure allows a separable solution of 
Helmholtz equation in cartesian coordinates.
The modal method consists in expanding the fields inside the corrugations in their own eigenfunctions 
(modes) that satisfy by themselves the boundary conditions at the sides of the cavities. In the
homogeneous region ($y \geq 0$), the total field $f$ is given by:
\begin{equation}
f(x,y)=e^{i(\alpha_0 x - \beta_0 y)} + \sum_{n=-\infty}^\infty 
R^q_n e^{i(\alpha_n x + \beta_n y)}\;\;,\;\;\;\;\;   \mbox{$q=s,p$}   \label{fred}
\end{equation}  
where $\alpha_0=k \sin \theta_0$, $\beta_0=k \cos \theta_0$,
\begin{equation}
\alpha_n=k \sin \theta_n = \alpha_0+ n \frac{2 \pi}{d}\;\;,
\end{equation}
\begin{equation}
\beta_n= \left\{ 
\begin{array}{ll}
\sqrt{k^2 -\alpha_n ^2} & \mbox{if $k^2\,  >\, \alpha_n^2$} \\ 
&  \\ 
i\,\sqrt{\alpha_n ^2-k^2 } & \mbox{if $k^2 \,<\, \alpha_n^2$}
\end{array}
\right. \;\;,  \label{betan}
\end{equation}
$k=2 \pi/\lambda$ is the wave number and $R^q_n$ is the unknown complex amplitude of the $n-th$ 
diffracted order . The superscript $q$ denotes the polarization case.

As mentioned above, the fields inside the corrugations are expressed in terms of infinite series.
Each zone $ij$ ($ij$ = 11, 12, 13 or 21) has its own expansion as follows:
\begin{equation}
f_{ij}(x,y)=\sum_{m=0}^\infty u_{ij m}^q(x)\, {\rm w}^q_{ij m}(y)\;\;,  \label{modal-ij}
\end{equation}
where 
\begin{equation}
u_{11 m}^q(x)= \left\{
\begin{array}{ll}
\sin{\left[\frac{m \pi}{a_{11}} x \right]} & \mbox{for $q=s$} \\
&  \\ 
\cos{\left[\frac{m \pi}{a_{11}} x \right]} & \mbox{for $q=p$}
\end{array}
\right.\;\;,       \label{u11q}
\end{equation}
\begin{equation}
u_{12 m}^q(x)= \left\{
\begin{array}{ll}
\sin{\left[\frac{m \pi}{a_{12}} (x-a_{11}) \right]} & \mbox{for $q=s$} \\
&  \\ 
\cos{\left[\frac{m \pi}{a_{12}} (x-a_{11}) \right]} & \mbox{for $q=p$}
\end{array}
\right.\;\;,        \label{u12q} 
\end{equation}
\begin{equation}
u_{13 m}^q(x)= \left\{
\begin{array}{ll}
\sin{\left[\frac{m \pi}{a_{11}} (x-(a_{11}+a_{12})) \right]} & \mbox{for $q=s$} \\
&  \\ 
\cos{\left[\frac{m \pi}{a_{11}} (x-(a_{11}+a_{12})) \right]} & \mbox{for $q=p$}
\end{array}
\right.\;\;,        \label{u13q} 
\end{equation}
\begin{equation}
u_{21 m}^q(x)= \left\{
\begin{array}{ll}
\sin{\left[\frac{m \pi}{a_{21}} x \right]} & \mbox{for $q=s$} \\
&  \\ 
\cos{\left[\frac{m \pi}{a_{21}} x \right]} & \mbox{for $q=p$}
\end{array}
\right.\;\;,       \label{u21q}
\end{equation}
\begin{equation}
{\rm w}_{11 m}^q(y)= \left\{
\begin{array}{ll}
C^s_{11m} \;\sin{[\mu_{11m} (y+h_1)]} & \mbox{for $s$ polarization} \\
&  \\ 
C^p_{11m} \;\cos{[\mu_{11m} (y+h_1)]} & \mbox{for $p$ polarization}
\end{array}
\right.\;\;,             \label{w11q}
\end{equation}
\begin{equation}
{\rm w}_{12 m}^q(y)=[A^q_{12m} \;\sin{(\mu_{12m} y)}+B^q_{12m} \; \cos{(\mu_{12m} y)}] 
\;\;\;\;\;\;\;\;\;\;\;\;\;\;\mbox{$q=s,p$}\;\;, \label{w12q}
\end{equation}
\begin{equation}
{\rm w}_{13 m}^q(y)= \left\{
\begin{array}{ll}
C^s_{13m} \;\sin{[\mu_{11m} (y+h_1)]} & \mbox{for $s$ polarization} \\
&  \\ 
C^p_{13m} \;\cos{[\mu_{11m} (y+h_1)]} & \mbox{for $p$ polarization}
\end{array}
\right.\;\;,             \label{w13q}
\end{equation}
\begin{equation}
{\rm w}_{21 m}^q(y)= \left\{
\begin{array}{ll}
C^s_{21m} \;\sin{[\mu_{21m} (y+h_1+h_2)]} & \mbox{for $s$ polarization} \\
&  \\ 
C^p_{21m} \;\cos{[\mu_{21m} (y+h_1+h_2)]} & \mbox{for $p$ polarization}
\end{array}
\right.\;\;,             \label{w21q}
\end{equation}
\begin{equation}
\mu_{ijm} = \left\{
\begin{array}{ll}
\sqrt{k^2 - \left[\frac{m \pi}{a_{ij}}\right]^2} & \mbox{if $k^2 \, >\,
\left[\frac{m \pi}{a_{ij}}\right]^2$} \\
&  \\ 
i\,\sqrt{\left[\frac{m \pi}{a_{ij}}\right]^2 - k^2} & \mbox{if $k^2 \, <\,
\left[\frac{m \pi}{a_{ij}}\right]^2$}
\end{array}
\right.\;\;,
\end{equation} 
and $C^q_{ijm}$, $A^q_{12m}$ and $B^q_{12m}$ are unknown complex 
amplitudes. Notice that $u_{ij 0}^s(x)=0$, and then the sum in eq. (\ref{modal-ij})
starts from $m=1$ in the $s$-case.
The functions $u_{11 m}^q(x)$ satisfy the appropriate boundary conditions
at $x=0$ and at $x=a_{11}$, the functions $u_{12 m}^q(x)$ satisfy the boundary
conditions at $x=a_{11}$ and at $x=a_{11}+a_{12}$, the functions $u_{13 m}^q(x)$ 
satisfy the boundary conditions at $x=a_{11}+a_{12}$ and at $x=d$, and the functions 
$u_{21 m}^q(x)$ satisfy the boundary conditions at $x=0$ and at $x=d$. On the other
hand, the functions ${\rm w}_{11 m}^q(y)$ and ${\rm w}_{13 m}^q(y)$ satisfy the boundary 
condition at $y=-h_1$, and the function ${\rm w}_{21 m}^q(y)$ satisfy the boundary
condition at $y=-(h_1+h_2)$, according to the case of polarization.

To solve the problem, the fields in each zone are matched imposing the boundary conditions
at the horizontal interfaces $y=-h_1$ and $y=0$. Then, expression (\ref{modal-ij}) with 
(\ref{u21q}) and (\ref{w21q}) is matched at $y=-h_1$ with the field in the central neck 
of the first layer (zone 12), given by (\ref{modal-ij}) with (\ref{u12q}) and (\ref{w12q}).  
For $0 \leq x \leq a_{11}$ and $a_{11}+a_{12} \leq x \leq d$, a null tangential electric field
is required. At $y=0$, the fields in zones 11, 12 and 13 are matched with those in $y \geq 0$,
given by eq. (\ref{fred}). All these conditions generate four $x$-dependent equations, that
are projected in appropriate bases (either the modal functions or the plane waves) to give
an infinite system of linear equations for the unknown amplitudes, for each polarization case. 
The explicit expressions of these equations can be found in the Appendix.
To find the numerical solution to this problem, we truncate the modal series in (\ref{modal-ij}) 
and the plane wave expansions in (\ref{fred}), and get a matrix system which is solved 
by a standard numerical technique of inversion.		

\section{Numerical results}

\begin{figure}[h]
\begin{center}
\includegraphics[width=4in]{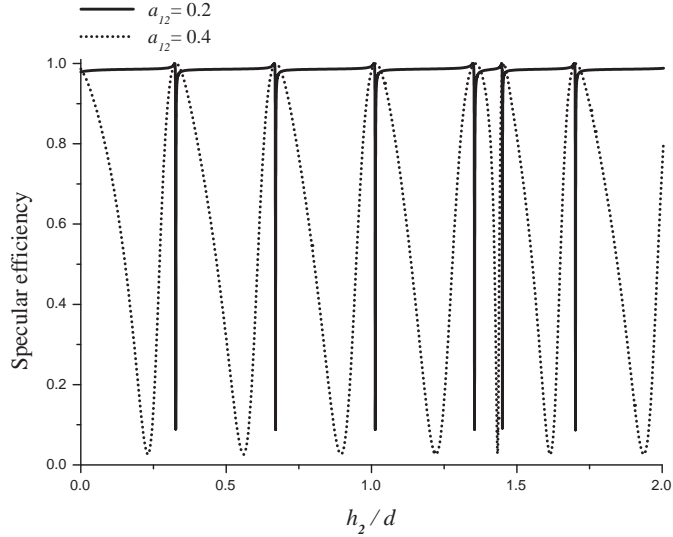}
\end{center}
\caption{Specular efficiency vs. $h_2/d$ for $s$ polarization. The grating parameters are $h_1/d=0.1$, 
$\lambda/d=0.65$, $\theta_0=0^\circ$, $a_{12}/d=0.2$ (solid)  
and $a_{12}/d=0.4$ (dashed)}
\end{figure}

In this section we analyze the diffraction by nested gratings for $s$ and $p$ polarization, 
paying particular attention to
the surface shape resonances that characterize their electromagnetic response. All the results
presented have been checked to satisfy energy conservation within an error less than $10^{-11}$.
The results for limit cases such as a comb grating ($h_2 \rightarrow 0$) \cite{DeSanto} and a 
bottle-shaped grating ($h_1 \rightarrow 0$) \cite{diana14} have also been verified. Even though the 
method was developed for an arbitrary angle of incidence,
in all the results presented we consider normal incidence. We analyze the grating response as a 
function of the depth of the bottom cavities $h_2/d$ and of the width of the neck $a_{12}/d$ (Figs. 2-6) and show the 
dependence of the behaviour of these structures as a function of the incident wavelength (Figs. 7-8).

In Fig. 2 we plot the specular efficiency versus $h_2/d$ for $s$ polarization, where: $h_1/d = 0.1$, 
$\lambda/d = 0.65$ and $a_{12}/d = 0.2$ (solid) and 0.4 (dashed). It can be noticed that there are minima in the efficiency curves for certain values of $h_2$. These minima are sharper and deeper for the narrower neck of the 
cavities, and can be associated with surface shape resonances. 

\begin{table}
\begin{center}
\begin{tabular}{|c|c|c|c|c|}
\hline
$m$ & $n$ & $h/d$ & $h_2/d$ & $pol.$\\
\hline
0 & 1 & 0.325 & 0.037 & $p$\\
1 & 1 & 0.344 & 0.32675 & $s$\\
2 & 2 & 0.427 & 0.3349 & $p$\\
0 & 2 & 0.650 & 0.468 & $p$\\
1 & 2 & 0.683 & 0.66985 & $s$\\
2 & 2 & 0.855 & 0.6628 & $p$\\
0 & 3 & 0.975 & 0.89 & $p$\\
1 & 3 & 1.030 & 1.01305 & $s$\\
2 & 3 & 1.283 & 0.9915 & $p$\\
0 & 4 & 1.300 & 1.29466 & $p$\\
1 & 4 & 1.374 & 1.35445 & $s$\\
3 & 1 & 1.462 & 1.45025 & $s$\\
0 & 5 & 1.625 & 1.339 & $p$\\
2 & 4 & 1.710 & 1.6342 & $p$\\
1 & 5 & 1.718 & 1.70125 & $s$\\
0 & 6 & 1.950 & 1.752 & $p$\\
%1.9624
\hline
\end{tabular}
\end{center}
\caption{Resonant depths $h/d$ for a perfectly conducting rectangular 
waveguide and resonant depths $h_2/d$ found in the nested structure for $a_{12}/d=0.2$, $h_1/d=0.1$, for a wavelength $\lambda/d=0.65$.}
\end{table} 

In Table 1 we compare the resonant depths of the cavities of the nested grating for $\lambda/d=0.65$ ($h_2$) with those corresponding to eigenmodes of a closed rectangular waveguide of side $d$, which are given by
\begin{equation}
\frac{h}{d}=\frac{n}{\sqrt{(2d/\lambda)^2-m^2}}\;\;,\mbox{$m,n \in Z_{\geq 0}$}\;\;.\label{modosguia}
\end{equation}
Notice that the first value of $m$ in (\ref{modosguia}) is 1 for $s$- polarization (the $x$- dependent part
of the electric field modes is a pure sine function, and $m=0$ would imply null field), whereas it is 0 for $p$-polarization (the $x$- dependent part of the magnetic field modes is a pure cosine function, and here $m=0$ implies uniform distribution in the $x$ direction), as it is explicitly shown in eq. (\ref{u21q}).
Besides, due to the symmetry imposed by the normally incident plane wave, only odd values of $m$ are
allowed for $s$-polarization, and only even values of $m$ for $p$-polarization. The number of decimal places
kept in the tables was determined by the resolution necessary to define correctly each minimum in the
efficiency curve. The sharper dips require more decimals than the smooth ones.

The resonant depths of the nested grating are close to those of the closed waveguide obtained by eq. 
(\ref{modosguia}), even though the present structure is open. The relationship between the width of the dip,
the quality of the resonance and the interior field can be better understood by inspection of the near field.

\begin{figure}[h]
\begin{center}
\includegraphics[width=5in]{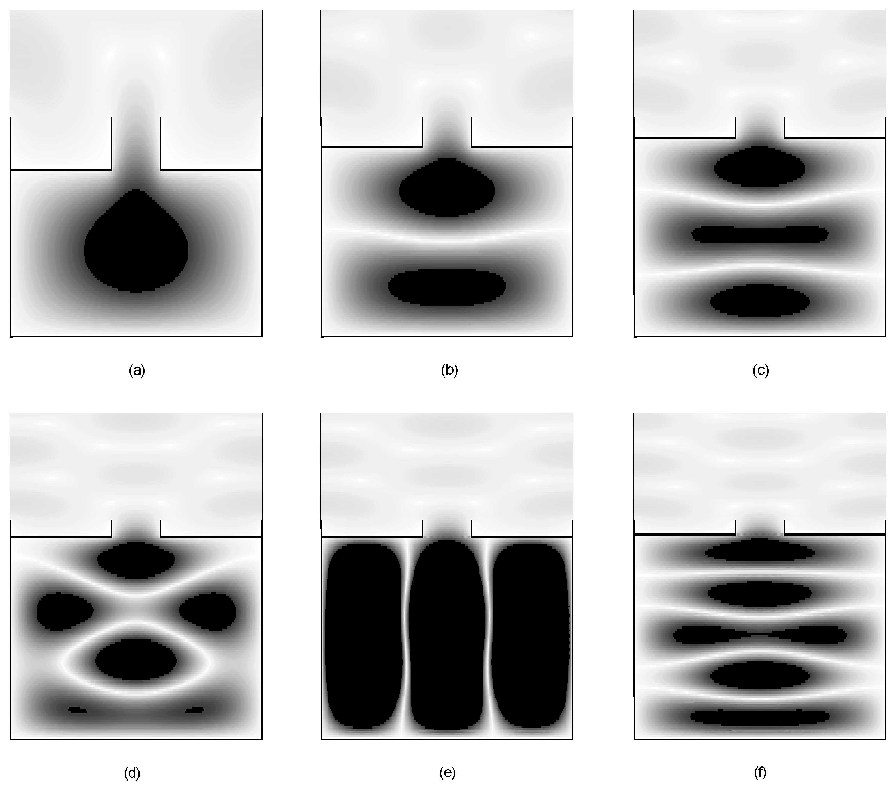}
\end{center}
\caption{Electric field intensity distribution for a grating with $h_1/d=0.1$, $a_{12}/d=0.2$,
$\lambda/d=0.65$, $\theta_0=0^\circ$ and $s$ polarization (solid line in Fig. 2): (a) $h_2/d=0.32675$; 
(b) $h_2/d=0.66985$; (c) $h_2/d=1.01305$; (d) $h_2/d=1.35445$; 
(e) $h_2/d=1.45025$; (f) $h_2/d=1.70125$.}
\end{figure}

\begin{figure}[h]
\begin{center}
\includegraphics[width=4in]{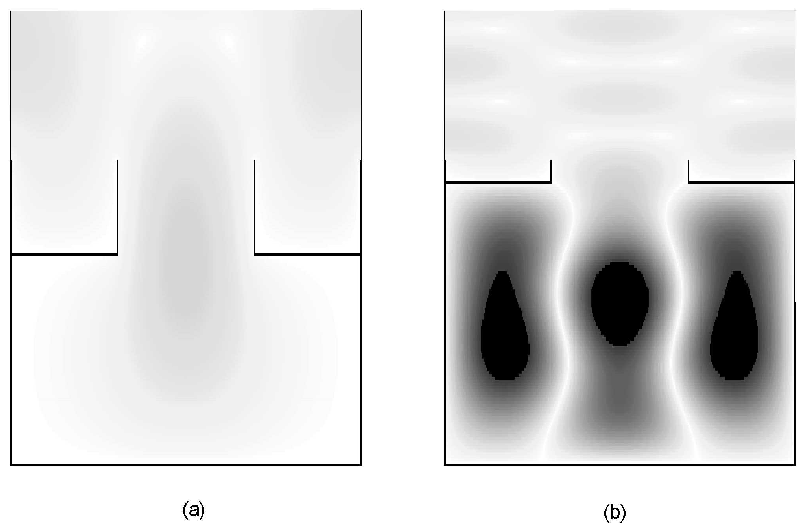}
\end{center}
\caption{Electric field intensity distribution for a grating with $h_1/d=0.1$, $a_{12}/d=0.4$,
$\lambda/d=0.65$, $\theta_0=0^\circ$ and $s$ polarization (dashed line in Fig. 2): (a) $h_2/d=0.23$; 
(b) $h_2/d=1.4335$.}
\end{figure}
 	    						
In Figs. 3 and 4 we plot the electric field magnitude (relative to the incident field) for different resonant depths of the structure considered in Fig. 2. In all the contour plots presented (Figs. 3, 4, 6 and 8) the black represents
maximum intensity, and the gray scale is maintained in all three figures.
Figs. 3a to 3f correspond to the resonant configurations for $a_{12}/d=0.2$, whereas Figs. 4a and 4b correspond to the $1^{st}$ and $5^{th}$ resonant cases for $a_{12}/d=0.4$. It can be observed in Fig. 3, that the contour plots of the inner field in the resonant configurations are similar to the configurations expected for the eigenmodes of a rectangular waveguide. Each one of the figures can be associated 
with a certain $mn$ mode of the closed waveguide: Fig. 3a. corresponds to the 11 mode, Fig. 3b to the 12, Fig. 3c 
to the 13, Fig. 3d to the 14, Fig. 3e to the 31, and Fig. 3f to the 15. Besides, the interior field is strongly 
intensified: the ratio between the maximum value inside and outside the structure varies between 6 (Fig. 3a) and 35 (Fig. 3e) for resonant configurations, whereas out of the resonance the field inside and outside the structure has roughly
the same value. Notice that the vertical scale is not maintained in all the figures, and therefore the depth of the cavities look equal although each one corresponds to a different depth.
When the neck of the cavities is widened, the quality of the resonances becomes lower, as observed
in the dashed curve of Fig. 2. At the same time, the resonant depths shift to lower values, thus moving further 
from the predicted depths for the closed waveguide. This behaviour is expected since this structure is less similar
to the closed one. 

The contour plots of electric field corresponding to this situation are shown in Fig. 4. Fig. 4a 
corresponds to the first dip ($h_2/d=0.23$) and Fig. 4b corresponds to the 5th. dip ($h_2/d=1.4335$). In the 
first case there is no intensification: the field takes the same values inside and outside the structure. The second case corresponds to the narrower dip of the dashed curve in Fig. 2, and consequently to the better
resonance for the range of $h_2$ considered. In this case there is an intensification, but it is significantly smaller than that of Fig. 3e (for the same resonant mode but $a_{12}/d=0.2$ instead of 0.4, and $h_2$ very close to that of 
the prevoius case). When the minimum of specular
efficiency is better localized, the enhancement of the field increases and so does the quality of the resonance.

\begin{figure}[h]
\begin{center}
\includegraphics[width=4in]{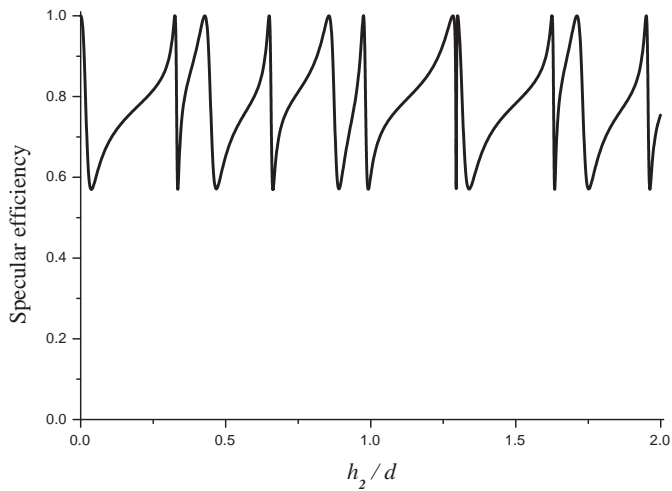}
\end{center}
\caption{Specular efficiency vs. $h_2/d$ for $p$ polarization. The grating parameters are $h_1/d=0.1$, 
$a_{12}/d=0.2$, $\lambda/d=0.65$, $\theta_0=0^\circ$.}
\end{figure}

The equivalent curve to the solid one in Fig. 2 but for $p$ polarization is shown in Fig. 5. It can be noticed that
the minima are not as deep and sharp as in the $s$-case, and this fact is also reflected in the near field plots 
of Fig. 6,
where we show the magnetic field for the first ($h_2/d=0.037$) and the 7th. ($h_2/d=1.29466$) minima of Fig. 5. 

\begin{figure}[h]
\begin{center}
\includegraphics[width=4in]{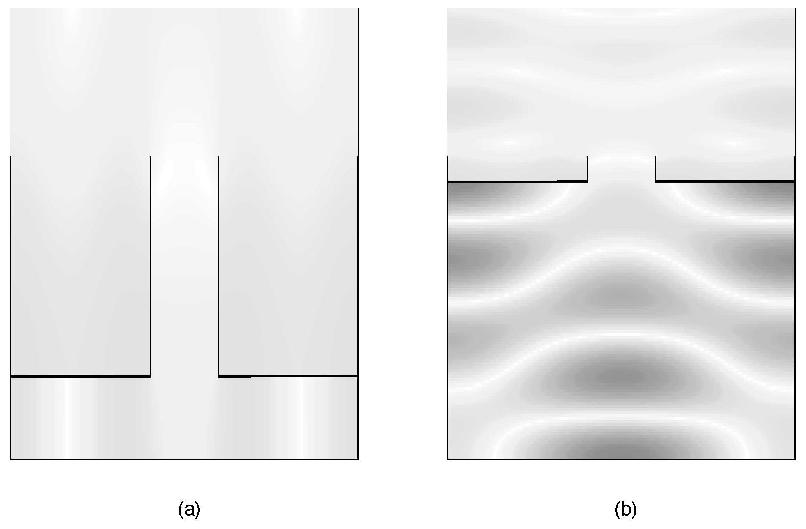}
\end{center}
\caption{Magnetic field intensity distribution for the grating and incidence conditions of Fig. 5: 
(a) $h_2/d=0.037$; (b) $h_2/d=1.29466$.}
\end{figure}

Not all the minima for the $p$-case are located close to those of the closed waveguide, listed in Table 1. This suggests that in general, these minima are not associated with resonances of the lower cavities in the nested grating, as it can be seen, for instance, in Fig. 6a, where we observe that the interior field is not intensified. However, there are certain depths that seem to 
fulfill the conditions for a resonant mode. This is the case of the 7th dip, whose corresponding magnetic
field plot is shown in Fig. 6b. According to Table 1, this dip is located very close to that of the waveguide for the
mode 04, and it can be clearly observed in the contour plot that the field distribution is that corresponding to
this mode: nearly uniform distribution in the $x$ direction and four half wavelengths in the $y$ direction.
Even though this is the most intense case, the intensification ratio is of about 6, roughly the same value of 
the less intensified mode of $s$ polarization. 

We have also analyzed the grating response as a function of the neck width $a_{12}$, and we have found that
for each configuration there is an optimum width for which the specular efficiency has its minimum (not shown). 
This value
arises from a compromise relationship between the narrowness of the neck and the necessary size of the aperture that allows the field to get into the cavity.

\begin{figure}[h]
\begin{center}
\includegraphics[width=4in]{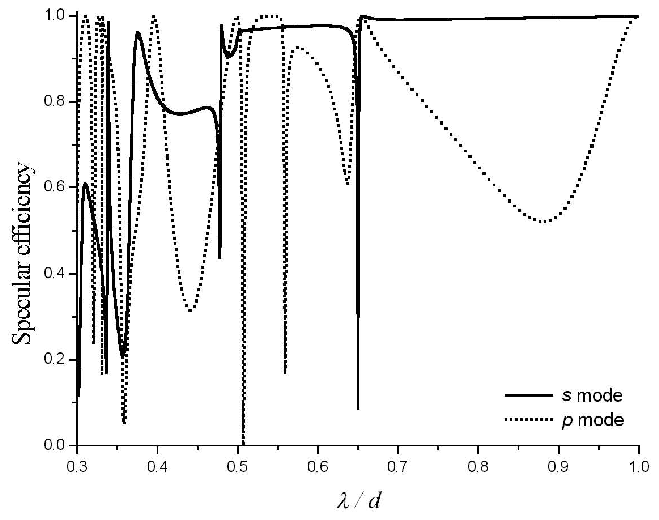}
\end{center}
\caption{Specular efficiency vs. $\lambda/d$ for a grating with $h_1/d=0.1$, $a_{12}/d=0.2$, 
$h_2/d=0.32675$, $\theta_0=0^\circ$, for $s$ (solid) and $p$ (dashed) polarization.}
\end{figure}

In Figs. 7-8 we explore the response of the nested structure as a function of the wavelength. The specular efficiency
for both polarization modes is shown in Fig. 7, for a grating with $h_1/d=0.1$, $h_2/d=0.32675$, $a_{12}/d=0.2$ and
$\theta_0=0^\circ$. For $\lambda/d$ greater than one, there is only one diffracted order, and the specular efficiency is
equal to unity (not shown). For smaller wavelengths, certain dips start to appear, that become more frequent as
$\lambda \rightarrow 0$. It is interesting to notice that there are certain wavelengths in which the efficiency is
nearly zero for one polarization, and at the same time it is maximum for the other. For instance, at $\lambda/d=0.65$
almost all the specularly reflected wave is $p$-polarized, whereas at $\lambda/d=0.5$, the reflected wave is $s$-polarized.
Then, this structure behaves as a polarizer for certain wavelengths, which can be varied by properly designing the 
cavities.

\begin{figure}[h]
\begin{center}
\includegraphics[width=6.5in]{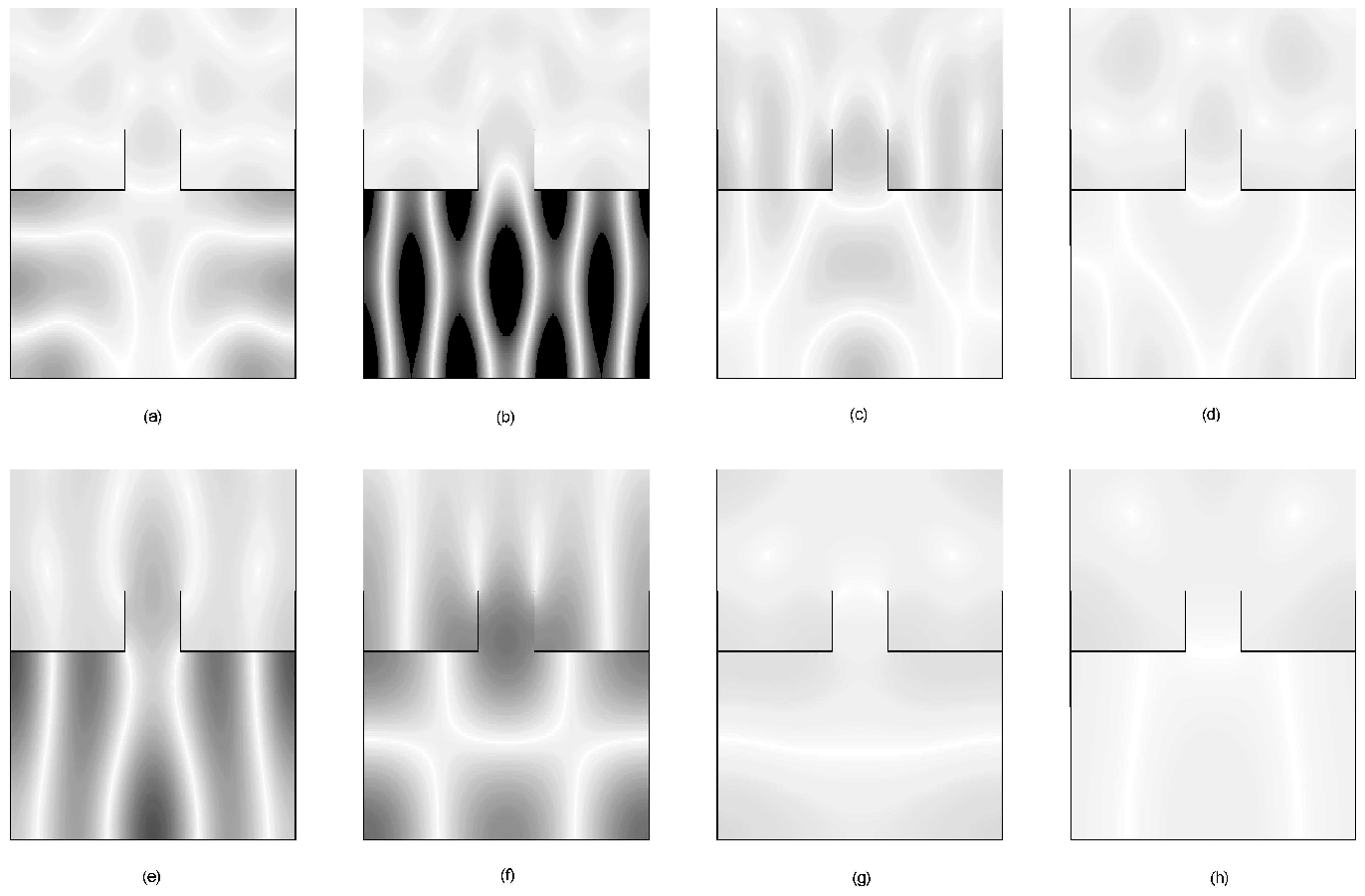}
\end{center}
\caption{Magnetic field intensity distribution for the grating of Fig.7, 
for $p$ polarization: (a) $\lambda/d=0.32099$; (b) $\lambda/d=0.331$; (c) $\lambda/d=0.35844$;
(d) $\lambda/d=0.44101$; (e) $\lambda/d=0.507$; (f) $\lambda/d=0.55895$; (g) $\lambda/d=0.63671$; 
(h) $\lambda/d=0.88201$.}
\end{figure}

Using eq. (\ref{modosguia}) to find the resonant wavelengths of a given waveguide, we
get the values listed in Table 2, where these values are compared with those corresponding to the dips of the nested
grating with $h_1/d=0.1$ and $a_{12}/d=0.2$. In the range of wavelengths considered, there are five dips for $s$ polarization and eight for $p$ polarization. As it can be noticed in the efficiency curve, some of them are sharper than others. This
suggests that their corresponding resonances are also better, in what concerns to the interior field intensification.
This can be confirmed by inspection of the interior field, shown as contour plots in Fig. 8, for $p$ polarization.
The figures are ordered by increasing resonant wavelength, and it can be observed that in most of the cases the magnetic field distribution is associated with a resonant mode $mn$ of the rectangular waveguide. For instance, Fig. 8a is the 02 mode, Fig. 8b is the 60 mode, Fig. 8e is the 40 mode, Fig. 8f is the 21 mode, etc. It is evident that the presence
of an aperture in the cavity modifies the field distribution of each mode, but most of them can still be identified.  
An interesting feature found in the wavelength dependence of the response, is the possibility of having modes
with $n=0$. These modes are not allowed in $s$ polarization since they imply null interior field; however, in the 
$p$-case, these modes represent an uniform distribution of the field in the $y$ direction. This is the case of the
modes 20, 40 and 60, shown in Figs. 8h, 8e and 8b, respectively. In general, the higher intensification is found when the resonant wavelengths are closer to those of the waveguide.
Another interesting phenomenon is the splitting of
the 41 mode: there are two neighbour dips close to the resonant wavelength predicted for the waveguide, but 
neither of them is too close (see Table 2). Even though the interior field configurations corresponding to those wavelengths (Figs. 8c and 8d) are different, both can be associated with the 41 mode, but none exhibits a significant intensification. The same effect appears in $s$ polarization for the 51 mode (the corresponding interior field plots
are not shown).

In general, the $s$-surface shape resonances in partially closed cavities are more significant than the $p$-ones, and this is in agreement with previous reports on multivalued structures \cite{diacla, diana13}.  

\begin{table}
\begin{center}
\begin{tabular}{|c|c|c|c|c|}
\hline
$m$ & $n$ & $\lambda_{w}/d$ & $\lambda_{g}/d$ & $pol.$\\
\hline
2 & 0 & 1.000 & 0.88 & $p$\\
0 & 1 & 0.653 & 0.63671 & $p$\\
1 & 1 & 0.621 & 0.650035 & $s$\\
2 & 1 & 0.547 & 0.55895 & $p$\\
4 & 0 & 0.500 & 0.507 & $p$\\
3 & 1 & 0.466 & 0.47762 & $s$\\
4 & 1 & 0.397 & 0.35844-0.44101 & $p$\\
5 & 1 & 0.341 & 0.33709-0.357 & $s$\\
6 & 0 & 0.333 & 0.331 & $p$\\
0 & 2 & 0.326 & 0.32099 & $p$\\
3 & 2 & 0.293 & 0.30156 & $s$\\
\hline
\end{tabular} 
\end{center}
\caption{Resonant wavelengths  for a perfectly conducting rectangular 
waveguide $\lambda_w/d$ and for the nested grating $\lambda_g/d$, for $h_2/d=0.32675$,
$a_{12}/d=0.2$ and $h_1/d=0.1$.}
\end{table}

\section{Conclusion}

The diffraction problem of a nested grating with rectangular cavities was solved for both polarization modes by means of the modal approach. The response of the grating has been analyzed, paying particular attention to the surface shape resonances. It was found that these resonances are stronger for $s$ than for $p$ polarization: the drops in the specular efficiency are sharper and deeper, and the electromagnetic field inside the cavities is significantly intensified. The resonant depths and the contour plots of the near field corresponding to the first modes of the structure are associated with the eigenmodes of a rectangular waveguide.

This study is a first approach to the analysis of the behaviour of nested structures in the presence of electromagnetic
waves. The advantages of using this kind of structures as parts of broadband antennas is now being analyzed, 
not only with the perfectly periodic model but also considering a finite structure and multilayers.

\section{Appendix}
By application of the boundary conditions at the horizontal interfaces, we
get four $x$-dependent equations that after appropiate projections generate an 
infinite system of linear equations for the unknown amplitudes. After a little
manipulation of the equations, and making the corresponding truncations, we can 
summarize the system for each polarization mode as follows. 

\begin{equation}
\overline{\overline{M}}^q \overline{A}^q = \overline{\overline{N}}^q \overline{B}^q\;\;,
\end{equation}
\begin{equation}
\overline{\overline{P}^q} \overline{A}^q = \overline{V}^q + \overline{\overline{Q}}^q \overline{{\cal R}}^q\;\;,
\end{equation}
\begin{equation}
\overline{\overline{S}}^q \overline{B}^q = \overline{W}^q + \overline{\overline{T}}^q \overline{{\cal R}}^q\;\;,
\end{equation}
where
\begin{equation}
M^q_{nk} = \left\{\begin{array}{ll}
			\delta_{nk} (1+\gamma_{12k}) + {\cal U}_{nk}^s (1-\gamma_{12k}) & \mbox{for $q=s$}\\
\\
			i [\delta_{nk} (1-\gamma_{12k}) + {\cal U}_{nk}^p (1+\gamma_{12k})] & \mbox{for $q=p$}
\end{array}
\right.
\end{equation}

\begin{equation}
N^q_{nk} = \left\{\begin{array}{ll}
			-i [\delta_{nk} (1-\gamma_{12k}) + {\cal U}_{nk}^s (1+\gamma_{12k})] & \mbox{for $q=s$}\\
\\
			\delta_{nk} (1+\gamma_{12k}) + {\cal U}_{nk}^p (1-\gamma_{12k}) & \mbox{for $q=p$}
\end{array}
\right.
\end{equation}

\begin{equation}
P^q_{nk} = \left\{\begin{array}{ll}
			\delta_{nk} & \mbox{for $q=s$}\\
\\
			\frac{2}{a_{21} \beta_n} e^{i(\mu_{12k} h_1-\alpha_n a_{11})} I^*(a_{12})_{kn} & \mbox{for $q=p$}
\end{array}
\right.
\end{equation}

\begin{equation}
Q^q_{nk} = \left\{\begin{array}{ll}
			\frac{\beta_k}{a_{12} \mu_{12n}} e^{i(\alpha_k a_{11} - \mu_{12n} h_1)} I(a_{12})_{nk} & \mbox{for $q=s$}\\
\\
			\delta_{nk} + \frac{2}{\beta_n a_{21} a_{11}} [1+e^{i(\alpha_k-\alpha_n)(a_{11}+a_{12})}] 
\sum_m \; \frac{\mu_{11m}}{(1+\delta_{m0})} \frac{(1-\gamma_{11m})}{(1+\gamma_{11m})} I(a_{11})_{mk} I^*(a_{11})_{mn}  & \mbox{for $q=p$}
\end{array}
\right.
\end{equation}

\begin{equation}
S^q_{nk} = \left\{\begin{array}{ll}
			\frac{2 i}{a_{21}} e^{i(\mu_{12k} h_1-\alpha_n a_{11})} I^*(a_{12})_{kn} & \mbox{for $q=s$}\\
\\
			\delta_{nk}& \mbox{for $q=p$}
\end{array}
\right.
\end{equation}

\begin{equation}
T^q_{nk} = \left\{\begin{array}{ll}
\delta_{nk} + \frac{2 \beta_k}{a_{11} a_{21}} [1+e^{i(\alpha_k-\alpha_n)(a_{11}+a_{12})}] 
\sum_m \;\frac{1}{\mu_{11m}} I(a_{11})_{mk} I^*(a_{11})_{mn}\frac{(1-\gamma_{11m})}{(1+\gamma_{11m})}  & \mbox{for $q=s$}\\
\\
			\frac{-i \mu_{12n}}{a_{12}(1+\delta_{n0})} e^{i(\alpha_k a_{11} - \mu_{12n} h_1)} I(a_{12})_{nk} & \mbox{for $q=p$}			
\end{array}
\right.
\end{equation} 

\begin{equation}
V^q_n = \left\{\begin{array}{ll}
-Q^s_{n0} & \mbox{for $q=s$}\\
\\
Q^p_{n0}-2 \delta_{n0} & \mbox{for $q=p$}
\end{array}
\right.
\end{equation}

\begin{equation}
W^q_n = \left\{\begin{array}{ll}
-T^s_{n0} +2 \delta_{n0}& \mbox{for $q=s$}\\
\\
T^p_{n0} & \mbox{for $q=p$}
\end{array}
\right.
\end{equation}
\begin{equation}
I(a)_{mk}=\left\{ \begin{array}{ll}
\int_{0}^a \sin\left(\frac{m \pi}{a} x\right) e^{i \alpha_k x} dx & \mbox{for $s$ polarization}\\
\\
\int_{0}^a \cos\left(\frac{m \pi}{a} x\right) e^{i \alpha_k x} dx & \mbox{for $p$ polarization}
\end{array}
\right.
\end{equation}
\begin{equation}
{\cal U}_{nk}^q=\left\{\begin{array}{ll}
\frac{4}{a_{12} a_{21} \mu_{12n}} \sum_{m}\; J_{km} J_{nm} \mu_{21m} \frac{1+\gamma_{21m}}{1-\gamma_{21m}} & \mbox{for $q=s$}\\
\\
\frac{4 \mu_{12n}}{a_{12} a_{21} (1+\delta_{n0})} \sum_{m}\; J_{km} J_{nm} \frac{1}{\mu_{21m} (1+\delta_{m0})} \frac{1+\gamma_{21m}}{1-\gamma_{21m}} & \mbox{for $q=p$}
\end{array}
\right.
\end{equation}

\begin{equation}
J_{mk}=\left\{ \begin{array}{ll}
\int_{0}^{a_{12}} \sin\left(\frac{m \pi}{a_{12}} x\right) \sin\left(\frac{k \pi}{a_{21}} (x+a_{11})\right) dx & \mbox{for $s$ polarization}\\
\\
\int_{0}^{a_{12}} \cos\left(\frac{m \pi}{a_{12}} x\right) \cos\left(\frac{k \pi}{a_{21}} (x+a_{11})\right) dx & \mbox{for $p$ polarization}
\end{array}
\right.
\end{equation}

\begin{equation}
\gamma_{ijm}=e^{2i \mu_{ijm} h_i} 
\end{equation}

In the above expressions the unknown modal amplitudes in the neck of the cavities have been redefined:
\begin{equation}
A^q_m=\left\{ \begin{array}{ll}
\frac{1}{2i} A^q_{12m} e^{-i \mu_{12m} h_1} & \mbox{for $q=s$}\\
\\
\frac{\mu_{12m}}{2i} A^q_{12m} e^{-i \mu_{12m} h_1}& \mbox{for $q=p$}
\end{array}
\right.
\end{equation}

\begin{equation}
B^q_m=\left\{ \begin{array}{ll}
\frac{1}{2i} B^q_{12m} e^{-i \mu_{12m} h_1} & \mbox{for $q=s$}\\
\\
\frac{\mu_{12m}}{2i} B^q_{12m} e^{-i \mu_{12m} h_1}& \mbox{for $q=p$}
\end{array}
\right.
\end{equation}
and ${\cal R}^q$ are the unknown Rayleigh coefficients.

\section*{Acknowledgments}

This work has been supported by Agencia Nacional de Promoci\'on Cient\'{\i}fica y 
Tecnol\'ogica (ANPCyT) under grant PICT98-4457, by CONICET (PEI 6216) and
by UBA (X150).

\end{document}